\documentclass{mn2e}

\usepackage{graphicx}
\usepackage{epstopdf}

\def\O2{O$_2~$}
\def\N2{N$_2~$}

\def\H2s{H$_2$S}
\def\SO2{SO$_2$~}
\def\water{H$_2$O~}

\def\H3OP{H$_3$O$^+$~}
\def\H2OP{H$_2$O$^+$~}

\def\ab0{$Ab_{0}~$}

\def\um{$\rm \mu m$~} 
\def\ltsim{\lower 0.7ex\hbox{$\buildrel < \over \sim\ $}}
\def\gtsim{\lower 0.7ex\hbox{$\buildrel > \over \sim\ $}} 
  
\def\cm3{cm$^{-3}~$}

\def\fn2{$F_{\rm N_2}~$}

\newcommand{\tm}{\mbox{$t_{\rm ice}$}}

\def\SC{Southern Coalsack}

\begin{document}

\title[Coalsack Chemical Chronology]{Chemical chronology of the
Southern Coalsack}

\author[S.D. Rodgers et al.]
{S.D. Rodgers$^1$\thanks{Email: rodgers@dusty.arc.nasa.gov}, S.B. Charnley$^1$, R.G. Smith$^2$ and  H.M. Butner$^3$ \\
 $^1$Space Science \& Astrobiology Division, MS
  245-3, NASA Ames Research Center, Moffett Field, CA 94035, USA
  \\ $^2$ School of Physical, Environmental \& Mathematical Sciences, The University of New
  South Wales, Australian Defence Force Academy, \\ Canberra ACT 2600,
  Australia \\ $^3$ Department of Physics,  James Madison University, 901 Carrier Drive, 
Harrisonburg, VA 22807-7702, USA \\ }

\maketitle

\begin{abstract}
We demonstrate how the observed H$_2$O ice column densities toward three
dense globules in the \SC\ could be used to constrain the ages of these
sources. We derive ages of $\sim 10^5$~yr, in agreement with dynamical
studies of these objects. We have modelled the chemical evolution of
the globules, and show how the molecular abundances are controlled by
both the gas density and the initial chemical conditions as the
globules formed. Based on our derived ages, we predict the column
densities of several species of interest. These predictions should be
straightforward to test by performing molecular line observations.
\end{abstract}


\section{Introduction}

Chemical reactions alter the composition of dense interstellar gas and
theoretical models of this evolution offer the prospect of determining
the relative ages of dense cores and molecular clouds (e.g., Stahler
1984; Leung, Herbst \& Huebner 1984).  The accuracy of these derived
ages will increase with that of the initial conditions of the model
calculations, i.e.  knowledge of the chemical state of the gas at some
arbitrary time in the past. In regions of active star formation the
prompt ($t$\ltsim 1yr) removal of icy grain mantles of known
composition, either by thermal evaporation or sputtering in shocks,
almost simultaneously alters the chemical composition of protostellar
cores and leads to a prolonged phase ($\sim 10^5$ years) of `hot core'
chemical evolution.  The destruction of abundant `parent' molecules
and the formation of `daughter' species in principle allows the time
since mantle removal to be measured from the presently observed gas
phase composition (Charnley, Tielens \& Millar 1992; Charnley 1997;
Hatchell et al. 1998).

Despite the fact that they are probably formed from more diffuse
material, of fairly simple chemical composition, estimating relative
`chemical ages' for molecular clouds and for dense (starless) cores
within those clouds is more problematic.  Models of quiescent dark
clouds have largely been based on so-called `pseudo-time-dependent'
models (e.g., Leung et al.\ 1984).  In this case, the physical
conditions remain fixed at dense cloud values throughout the
calculation whereas the initial conditions for the heavy elements are
those expected from the diffuse interstellar medium. The chemical
evolution in these purely gas phase models can be roughly divided
between the abundances present at `early-times', prior to a major
fraction of the gas phase carbon being incorporated into CO ($\sim
10^5$ years), and `late-times', approximately when the chemistry has
reached a steady-state ($\sim 10^6-10^7$ years).  Time-dependent
models of dark clouds have shown that the carbon-chain molecules are
best explained by `early-time' chemistry whereas several other
molecular abundances, such as $\rm N_2H^+$ and $\rm NH_3$, are closer
to the predicted steady-state values. This has led to the idea that
observed spatial compositional gradients (e.g.\ involving the
carbon-chains and ammonia) are due to a differing chemical ages in
individual clouds (e.g.\ Hirahara et al.\ 1992; Taylor, Morata \&
Williams 1998).

Much effort has been made by theoretical astrochemists to reconcile
the apparent chemical youth of molecular clouds with other evidence
which suggests that these objects are long-lived entities (e.g., Nejad, Williams \& Charnley 1990).
For example, in the `classical' picture of molecular cloud evolution and low-mass
star formation (Shu, Adams \& Lizano 1987), the ambipolar diffusion
time-scale of a plasma with fractional ionization $x_ {\rm e }$ 
\begin{equation}
{ t_{\rm AD} ~=~ 7.3 \times 10^{6}  \left({x_ {\rm e } \over 10^{-7}
} \right)\rm ~~ yr }
\end{equation} 
(McKee et al.\ 1993) governs the loss of magnetic support which is a
prerequisite for gravitational collapse (Shu et al.\ 1987).
In this picture, molecular clouds are inferred to have lifetimes of $\sim 10$~Myr 
(Mouschovias, Tassis \& Kunz 2006).
However,
recent studies suggest that  molecular cloud
formation, dissipation and star formation, are all rapid dynamical
processes (Elmegreen 2000; Hartmann, Ballasteros-Paredes \& Bergin
2001; Mac Low \& Klessen 2004). In this revised picture, clouds form through
the dissipation of supersonic turbulence and the important dynamical
time-scale is the gravitational free-fall time at number density $n_{\rm H}$
\begin{equation}
{ t_{\rm ff} ~=~ { { 4.35 \times 10^{7} }  { n _{\rm H}^{-1/2} } }
 \rm ~~~ yr }
\end{equation}
Although the debate over the mode of star formation is not yet settled
(e.g., Mouschovias et al.\ 2006; Ballesteros-Paredes \& Hartmann 2006),
if dense clouds and stars
form within a few free-fall times, with the cloud life-time being
$\sim$ 1--3 Myr (Hartmann et al.\ 2001), determining the chemical ages
of molecular clouds may, after all, only require consideration of the
abundances during the `early time' phase.

To test and refine the idea that chemical models could be used to
measure the (relative) ages of molecular clouds requires two
additional criteria.  First, we would like to model the chemical
evolution in a molecular cloud for which there is independent evidence
that it formed recently.  The Southern Coalsack is a well-studied
region which fulfills this criterion. The Coalsack is a group
of molecular globules (or clumps) where the low average gas densities and the
lack of discernible star formation activity (Nyman, Bronfman \&
Thaddeus 1989; Kato et al.\ 1999), the presence of a diffuse
inter-clump medium, as evidenced by CH$^+$ observations (Centuri\'on,
C\`assola \& Vladilo 1995), and the apparent dynamical youth of the
major globule (Lada et al.\ 2004), all suggest that it is a
region where simplistic chemical models may actually be applicable.
Second, we would like to use another time-scale to constrain the
chemical evolution.  The time-scale over which icy grain mantles are
laid down in molecular clouds provides a natural constraint that is
independent of the precise chemical state of the gas (Charnley,
Whittet \& Williams 1990). This time-scale, \tm, is determined
by the rate at which oxygen atoms collide with and stick to dust grains,
and depends on the gas density, the atomic oxygen abundance, 
and the total effective grain surface area 
(see $\S$\ref{sec:tm}).

Ice mantles have been detected towards
several globules in the Southern Coalsack by Smith et al.\ (2002) and
this allows us to estimate the mantle formation time-scale for each.
In this paper, we use the mantle formation time-scale as the
basic clock by which to measure the chemical chronology of dense
globules in the \SC, and thereby predict their relative chemical
compositions.

\section{The Southern Coalsack}

The Southern Coalsack is a conspicuous dark molecular cloud complex in
the southern sky. Unlike similar regions in the northern Milky Way,
such as Taurus, Ophiuchius and Perseus, a striking feature of the
\SC\ is the, apparently complete, absence of star formation (Nyman et
al. 1989; Kato et al. 1999; Bourke, Hyland \& Robinson 1995; Racca,
G\'omez \& Kenyon 2002). Optical and millimetre-wave observations of
molecules towards the background field star HD 110432, which
sample a line-of-sight through the inter-clump medium,
show that the \SC\ is pervaded by
translucent ($ A_{ V}\approx 1-3$ mag.), diffuse ($n_{\rm H} \sim
10^2\rm cm^{-3} $) material (Codina et al.\ 1984; van Dishoeck \&
Black 1989; Gredel, van Dishoeck \& Black 1994; Centuri\'on et
al.\ 1995; Rachford et al.\ 2001). The low CO abundance (Codina et
al.\ 1984), and the presence of only simple molecules, are consistent
with the chemistry expected in translucent clouds.

Tapia (1973) identified three dark globules (hereafter Globules 1,2 \&
3) which showed higher than average extinction. Apart from isotopomers
of CO, the \SC\ has thus far only been mapped in OH, H$_2$CO, and H~I
(Brooks, Sinclair \& Manefield 1976; McClure-Griffiths et
al.\ 2001). Nyman et al.\ (1989) mapped the \SC\ in CO lines and
showed that Tapia's globules were also peaks in CO
emission. Observations of $^{13}$CO and C$^{18}$O have identified
several other distinct cores (Vilas-Boas, Myers \& Fuller 1994; Kato
et al.\ 1999).  These studies indicate that dense condensations in the
Coalsack have densities in the range $n_{\rm H} \sim 10^3-10^4~\rm
cm^{-3} $ with gas and dust temperatures of around 10~K (Jones, Hyland
\& Bailey 1984).  An infrared extinction study of Globule 2 (Lada et
al.\ 2004) indicates that it is in a state of dynamical equilibrium,
and suggest that it formed only recently. Thus, the Southern Coalsack
is a region where dense cloud cores have recently condensed out of a
more diffuse, low-extinction, interclump medium. 

Smith et al.\ (2002)
performed 3\um spectroscopy towards eight field
stars, located behind the Southern Coalsack, to search for \water ice
absorption towards several of the dense globules.  Six of these
sources lay behind or near Globule 2 and one behind each of the other
two.  Smith et al.\ detected water ice in each of Tapia's globules with
an additional four tentative detections in Globule 2. 
The largest ice column densities were found toward
stars with the largest visual extinctions, as found in many other
interstellar clouds (e.g., Whittett et al.\ 1988). In general, the fact that the water
features are only apparent above a certain level of $A_V$ is not 
direct proof that the ice must be present in the densest regions,
although it is suggestive that this is the case. In the Coalsack, however,
the fact that Smith et al.'s field stars lay behind the globules, and we
know from the CO line maps that these are regions of higher density,
leads us to conclude that the observed H$_2$O ice is frozen out in 
the cold, dense globules, and not in the inter-clump medium.
In what follows, we use data from Smith et al.\ for their sources designated
SS1-2, SS2-2 and SS3-1 to derive a mantle formation time-scale in each
of Tapia's Globules 1, 2 \& 3.

\section{Ice formation time-scales}
\label{sec:tm}

We now show how observations of the ice column density can be used to
estimate the ice mantle formation time-scales.  Accretion and
hydrogenation of oxygen atoms will lead to a spherical (silicate)
grain core of radius $a$ being covered by a water ice mantle, and the
mean particle radius growing as
\begin{equation}
{ r(a,t)  ~ = ~ a ~ + ~ d(a,t)  }
\end{equation}
where $d(a,t)$ is the instantaneous mean mantle thickness.
A spherical grain of instantaneous radius $r(a,t)$ increases in size
 by accretion of mass through oxygen atoms sticking to the surface at
 a rate (Wickramasinghe 1967)
\begin{equation}
{dr \over dt} ~ = ~ {S  \over \rho_{\rm ice} } 
\left ( {kT m_{\rm O}\over 2 \pi } \right)^{1/2}  n_{\rm O}(t) 
\label{eqn:drdt}
\end{equation}
where $T$ is the gas temperature, $S$ is the sticking efficiency,
$m_{\rm O}$ is the mass of an oxygen atom, $\rho_{\rm ice}$ is the
bulk density of water ice, and $n_{\rm O}(t)$ is the number density of
O atoms in the gas.  For simplicity, we have ignored the increase in
mass due to the accretion of two hydrogen atoms per oxygen.  Because
both the molecular accretion rate and the number of molecules required
to build up one monolayer of coverage are proportional to the surface
area of the grain, the radial growth of grains is independent of grain
size. Hence, at any particular time all grains will be covered by the
{\rm same} thickness of ice.


An  ice mantle will contain on average $W(a,t)$ water molecules
\begin{equation}
 {W(a,t) ~ = ~ {4 \pi \rho_{\rm ice} \over 54 m_{\rm H}  }  
\lbrack r^3(a,t) ~-~a^3\rbrack   } 
\end{equation}
where $m_{\rm H}$ is the hydrogen nucleon mass. A pathlength $L$ through a molecular
cloud with uniform dust and gas number densities, $n_{\rm gr}(a)$ and
$n_{\rm H}$, has an ice column density of
\begin{equation}
{ N_{\rm ice}(t)~ = ~ \int_0^L \int_{a_-}^{a_+}  W(a,t) {n_{\rm gr}(a)}~da~ dl 
  }
\end{equation}
If the hydrogen column density,
$N_{\rm H}$ ($=n_{\rm H}L$), is related to the visual extinction
through  
\begin{equation}
{ N_{\rm H}  ~ = ~ 1.6 \times 10^{21}   A_{ V} \rm ~~cm^{-2} }
\end{equation}
(Bohlin, Savage \& Drake 1978) then the time to accumulate this much ice, the mantle formation
time-scale, \tm, satisfies
\begin{equation}\label{eqn:intm}
{  \int_{a_-}^{a_+} \lbrack r^3(a,\tm) ~-~a^3\rbrack a^{-3.5}~da  ~ = ~
5.8 \times 10^{-20}{N_{\rm ice}(\tm) \over \rho_{\rm ice} A_{ V}}
}
\end{equation}
Here we have  assumed that the dust obeys an MRN size distribution
(Mathis, Rumpl \& Nordsieck 1977) such that
\begin{equation}
{ { n_{\rm gr}(a)  }    ~ = ~ A  a^{-3.5} n_{\rm H} }
\end{equation}
in a range bounded by ${a_-}= 0.005 ~\mu \rm m$ and ${a_+}= 0.25 ~\mu
\rm m$, and where $A = 7.76 \times 10^{-26} \rm cm^{2.5}/H $ for
silicate dust (Draine \& Lee 1984).  The ice mantle thickness at this
time, $d_{\rm m}=d(\tm)$, is just $r(a,\tm) - a $, and so we can
evaulate the integral in (\ref{eqn:intm}) to obtain a cubic equation
in $d_{\rm m}$. As $N_{\rm ice}$/$A_V$ is determined observationally,
the cubic can be solved to obtain $d_{\rm m}$ for any source.
  
We must now calculate the time, \tm, required to accrete this
thickness of ice.  Accretion on to grains removes oxygen atoms from the gas at a rate
\begin{equation}\label{eqn:dndt}
{dn_{\rm O} \over dt} ~ = ~ - \frac{\Lambda_{\rm
gr} (t)}{4} n_{\rm H} S \left ( {8kT }\over \pi m_{\rm O} \right)^{1/2} n_{\rm O}(t)
\end{equation}
where the total available grain surface area per nucleon at time $t$, $\Lambda_{\rm
gr} (t)$, is
\begin{eqnarray}
\lefteqn{\Lambda_{\rm gr} (t) ~=~ \left \langle 4 \pi r^2(a,t)n_{\rm
gr}(a)/n_{\rm H}\right \rangle } \nonumber \\
&~~ ~=~ & \int_{a_-}^{a_+} 4 \pi
r^2(a,t)n_{\rm gr}(a)/n_{\rm H} ~da \nonumber \\
& ~~~=~ & 9.75 \times 10^{-25} \int_{a_-}^{a_+}r^2(a,t) a^{-3.5}~da
\end{eqnarray}
%
%
The factor of 4 in eqn.\ (\ref{eqn:dndt}) is due to the fact that $\Lambda_{\rm gr}$
represents the total grain surface area ($4\pi r^2$), whereas the cross-section for accretion
onto each grain is equal to $\pi r^2$.

Subject to some approximations, an analytical solution for \tm\ can be
obtained. Firstly, accretion on to grains is the dominant loss route
for O atoms in dark clouds, so we neglect the effects of other
chemical processes on $n_{\rm O}(t)$.  We also assume that the
underlying power-law for $n_{\rm gr}(a)$ is not modified by
coagulation.  For time-scales sufficiently short that $\Lambda_{\rm
  gr} (t)$ is approximately unchanged (i.e.\ only thin mantles have
formed), the total surface area of the bare grains integrated over the
MRN size distribution, $\Lambda_{\rm gr} (0)$, is equal to
$2.37\times10^{-21}$ cm$^2$/H\@.
For constant $\Lambda_{\rm gr}(0)$, the solution of eqn.\ (\ref{eqn:dndt}) is 
\begin{equation}
  n_{\rm O}(t)  = n_{\rm O}(0) e^{-t/t_{\rm acc}}
\end{equation}
where $n_{\rm O}(0)$ is the initial gas-phase oxygen atom abundance
and $t_{\rm acc}$ is their accretion time-scale
\begin{equation}
 t_{\rm acc} = \frac{1}{\Lambda_{\rm gr}(0) n_{\rm H} S} \left(
  \frac{2\pi m_{\rm O}}{kT} \right)^{1/2} 
\end{equation} 
Integration of eqn.\ (\ref{eqn:drdt}) then yields
\begin{equation}\label{eqn:ddmax}
   d(t) = d_{\rm max} \left[ 1 - e^{-t/t_{\rm acc}} \right]
\end{equation}
where $d_{\rm max} = (18m_{\rm H}n_{\rm O}(0)/\Lambda_{\rm gr}(0) n_{\rm
 H}\rho_{\rm ice})$ is the maximum thickness of ice that can be
attained when all the oxygen has frozen out. 
With $d_{\rm m} $ set from the observational data through equation  (\ref{eqn:intm}),  the mantle formation time is found from  
\begin{equation}\label{eqn:tm1}
  \tm = t_{\rm acc}\, \ln \left[ 1 - \frac{d_{\rm m}}{d_{\rm max}}\right]^{-1}
\end{equation}

In order to calculate \tm\ in the Coalsack, we adopt a sticking coefficient
of unity, and assume a gas and dust temperature of 10~K (Jones
et al.\ 1984).
Experimental measurements of amorphous water ice
deposited at low temperatures indicate densities in the range
0.6--1~g~cm$^{-3}$, depending on the details of the ice deposition
(e.g.\ Jenniskens \& Blake 1994; Stevenson et al.\ 1999). Therefore,
we adopt $\rho_{\rm ice} = 0.75$~g~cm$^{-3}$.  Values
of $N_{\rm ice}$, $A_V$, and $n_{\rm H}$ for each of the three
globules are taken from the observations of Kato et al.\ (1999) and
Smith et al.\ (2002). The only remaining parameter required to
calculate \tm\ is the initial oxygen atom
abundance in the gas, which depends on the elemental abundances in the
Coalsack, and on the original chemical conditions in each globule
(i.e.\ how much oxygen was locked up in CO and other molecules). For
the chemical models described in the following section, O atom
abundances of $n_{\rm O}(0)/n_{\rm H} = 1.6\times10^{-4}$ and $2.9
\times 10^{-4}$ are assumed, depending on whether the chemistry in
each globule was initially appropriate for translucent (T) or diffuse
(D) material.

Table~1 lists the physical conditions observed in each globule, and
the corresponding values of $d_{\rm m}$ and \tm\ derived for each
object. In all cases the derived mantle thicknesses are
significantly less than the maximum possible value, $d_{\rm max}$,
showing that only a small fraction of the total oxygen is frozen out
as water, and that each globule must be much significantly younger
than the accretion time-scale. The \tm\ values derived are consistent with these
globules having formed recently (cf.\ $\sim 2 \times 10^{5}$ years;
Lada et al.\ 2004).
The principal uncertainties in our derived values of \tm\
come from uncertainties in the values of  $\Lambda_{\rm gr}(0)$ and the
initial gas-phase oxygen atom abundance. For example, if we truncate the
grain size distribution at $a_- = 100$~\AA\ instead of 50~\AA, this reduces the
effective grain surface area by a factor of almost two-thirds, and thus increases
the derived values of \tm\ by 50 per cent. Similarly, because the initial values
for $n_{\rm O}$ differ by a factor of two between models D and T, the derived
ages differ by a similar factor. Hence, determining the absolute age of each core
requires knowledge of the initial chemical conditions and the grain size distribution.
However, it is likely that each globule coalesced from material with similar
properties, and so the {\it relative} ages will be much more accurate.


\begin{table*}
\caption{Globule properties and ice mantle formation time-scales in the Coalsack}
\label{tab:globs}
\begin{center}
\begin{tabular}{crrrrrrr}
\hline\hline 
Globule & $N_{\rm ice}^\dagger$~~~~~ & $A_{\rm V}^\dagger$~~ & $n_{\rm H}^\ddagger$~~~~~
   & x(\rm CO)$^\ddagger$  & $d_{\rm m}$~  & \tm(D)~~ & \tm(T)~~
\\
 & $(10^{17}~\rm cm^{-2})$ & (mag.) & $(10^3~\rm cm^{-3})$ & $(10^{-5})$ &
 (\AA)~ &  ($10^5~\rm yr$) & ($10^5~\rm yr$)
\\ \hline
   1  &  9.1~~~~~~  &  11.2~~  &  4.0~~~~~  &   7.5~~  & 85~ & 2.0~~~ & 3.7~~~ \\
   2  &  5.7~~~~~~  &  13.3~~  &  8.2~~~~~  &  13.7~~  &  45~ & 0.5~~~ & 1.0~~~ \\
   3  &  2.0~~~~~~  &   9.9~~  &  5.4~~~~~  &   8.6~~  &  21~  & 0.4~~~ & 0.7~~~ \\
\hline
\end{tabular} \\
\end{center}
\begin{flushleft}
\footnotesize
\noindent
 $^\dagger$ Smith et al. (2002).
 $^\ddagger$ Kato et al. (1999); for globule 3 we have used values quoted for their core 4.
 \tm(D) and \tm(T) are the ice formation time-scales corresponding to
 different values of the initial O atom abundance (see text).
\end{flushleft}
\end{table*}


\section{Chemical model}

We now want to use the derived \tm\ values to find chemical
discriminants of the different ages. The values of \tm\ measure
the time since the ice mantles began to be laid down, and it is well
known that mantle formation does not occur until the visual extinction
reaches a critical threshold of $A_{\rm V} \sim
3$~mag.\ (e.g.\ Whittet et al.\ 1988). In the Coalsack, Smith et
al.\ (2002) derive an extinction threshold in the range 2.6--7.6
mag. Therefore, the appropriate initial conditions for the chemical
model at $t = 0$ are determined by the chemical state of the cloud
when $A_{\rm V} \approx 3$--8~mag., which are in turn determined by
the physical evolution of the cloud from the diffuse phase to a dense
core. In the limit that core formation occured rapidly, the initial
abundances will be identical to those in the diffuse ISM,
i.e.\ principally atomic oxygen and nitrogen with carbon and sulphur
in ionic form. On the other hand, if the cloud evolved slowly, so that
it spent a long time in the `translucent' phase ($1 \ltsim A_{\rm V}
\ltsim 3$~mag.), then the cloud material will have been mainly
molecular (CO and N$_2$, with only O atomic) when $A_{\rm V}$ reached
the critical value and ices began to form. Therefore, we have considered  two
models corresponding to initally diffuse and translucent chemical
abundances: the abundances at $t = 0$ for both models are listed in
table~\ref{tab:init}.
 
The gas-phase elemental abundances are taken from Savage \& Sembach
(1996), except for sulphur which is assumed to be significantly
depleted relative to the diffuse ISM (Ruffle et al.\ 1999). In our `standard'
model runs we also
assume complete depletion of heavy elements, although
the effects of non-zero metal abundances are discussed below in $\S$\ref{sec:metals}.
The chemical
reaction network is described in Rodgers \& Charnley (2001), and has
been updated to incorporate several recently measured dissociative
recombination channels (e.g.\ Geppert et al.\ 2006). All neutral
species except H$_2$ and He freeze out onto grain mantles with
a sticking coefficient of unity. We adopt a temperature of 10~K and a
cosmic ray ionization rate of $3\times10^{-17}$~s$^{-1}$. Values for
$n_{\rm H}$ and $A_{\rm V}$ are set equal to those derived for each of
the three globules (table~\ref{tab:globs}).

\begin{table}
\caption{Initial abundances relative to H$_2$}
\label{tab:init}
\begin{center}
\begin{tabular}{lll}
\hline\hline
Species & \multicolumn{2}{c}{Abundance} \\
        & Model D & Model T \\
\hline
 C$^+$  &  2.8(-4) & 2(-8) \\
 C      &  0       & 2(-5) \\
 CO     &  0       & 2.6(-4) \\
 O      &  5.8(-4) & 3.2(-4) \\
 N      &  1.6(-4) & 0 \\
 N$_2$  &  0       & 8(-5) \\
 S$^+$  &  1(-7)   & 1(-7) \\
 He     &  0.2 & 0.2 \\ 
\hline\hline
\end{tabular}\\
\end{center}
{\small Model D corresponds to the abundances expected if the globules
  condense rapidly from diffuse cloud material. In model T the initial
  conditions represent those in low-density translucent ($A_{\rm V}
  \approx 2$) gas. $a(-b)$ represents $a\times10^{-b}$.}
\end{table}
  
%
\section{Results}

\begin{figure*}
\includegraphics[width=6in]{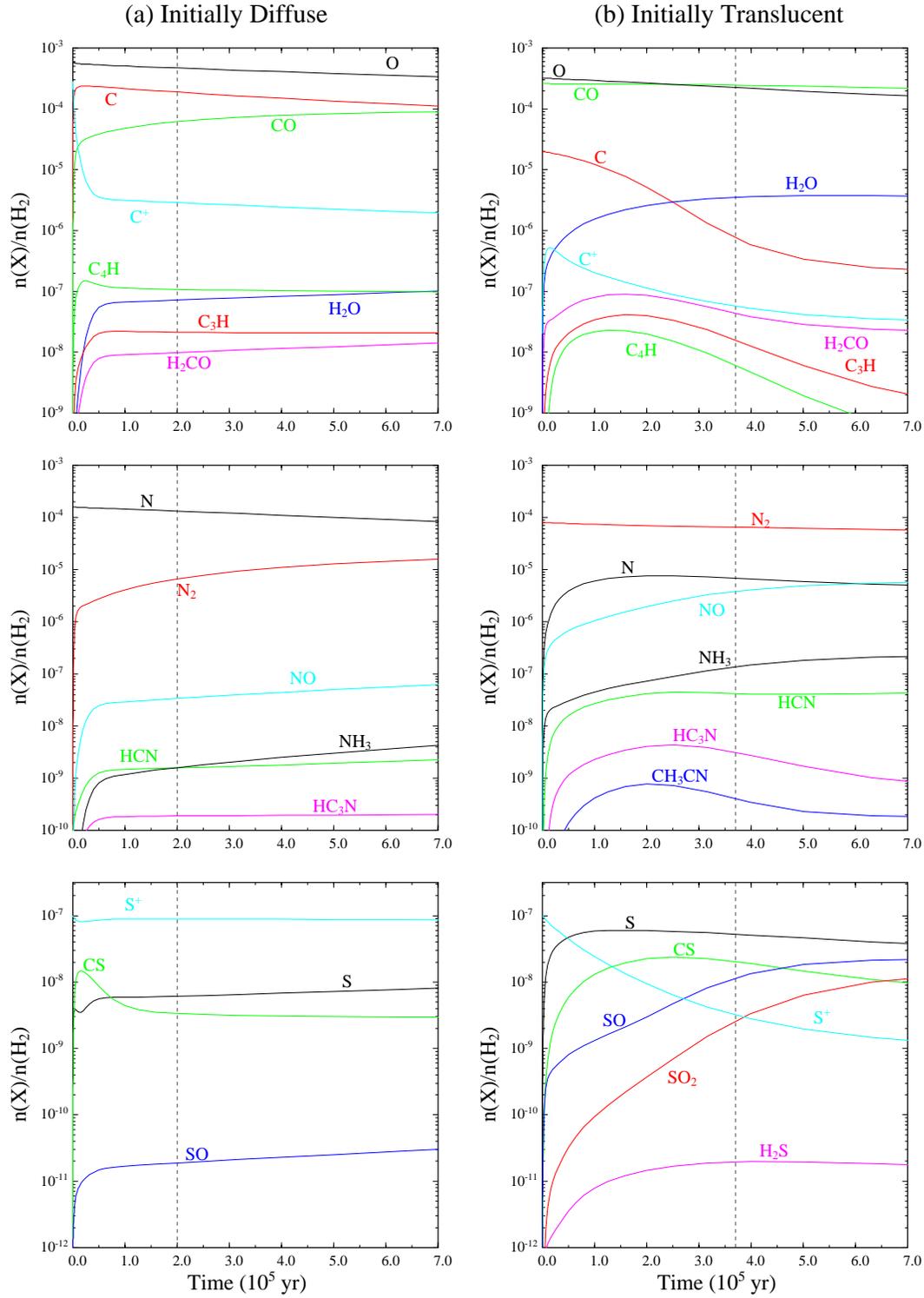}
\caption{Chemical evolution in Coalsack Globule 1. (a) The left panels
  show the model results for initially diffuse cloud abundances,
  and (b) the right hand panels show the evolution from initially
  transclucent cloud conditions. The gray vertical dashed lines
  mark the ice mantle formation time-scale, \tm. The time-scale is
  longer in (b) because the initial abundance of atomic oxygen is
  smaller.}
\label{fig:G1}
\end{figure*}

\begin{figure*}
\includegraphics[width=6in]{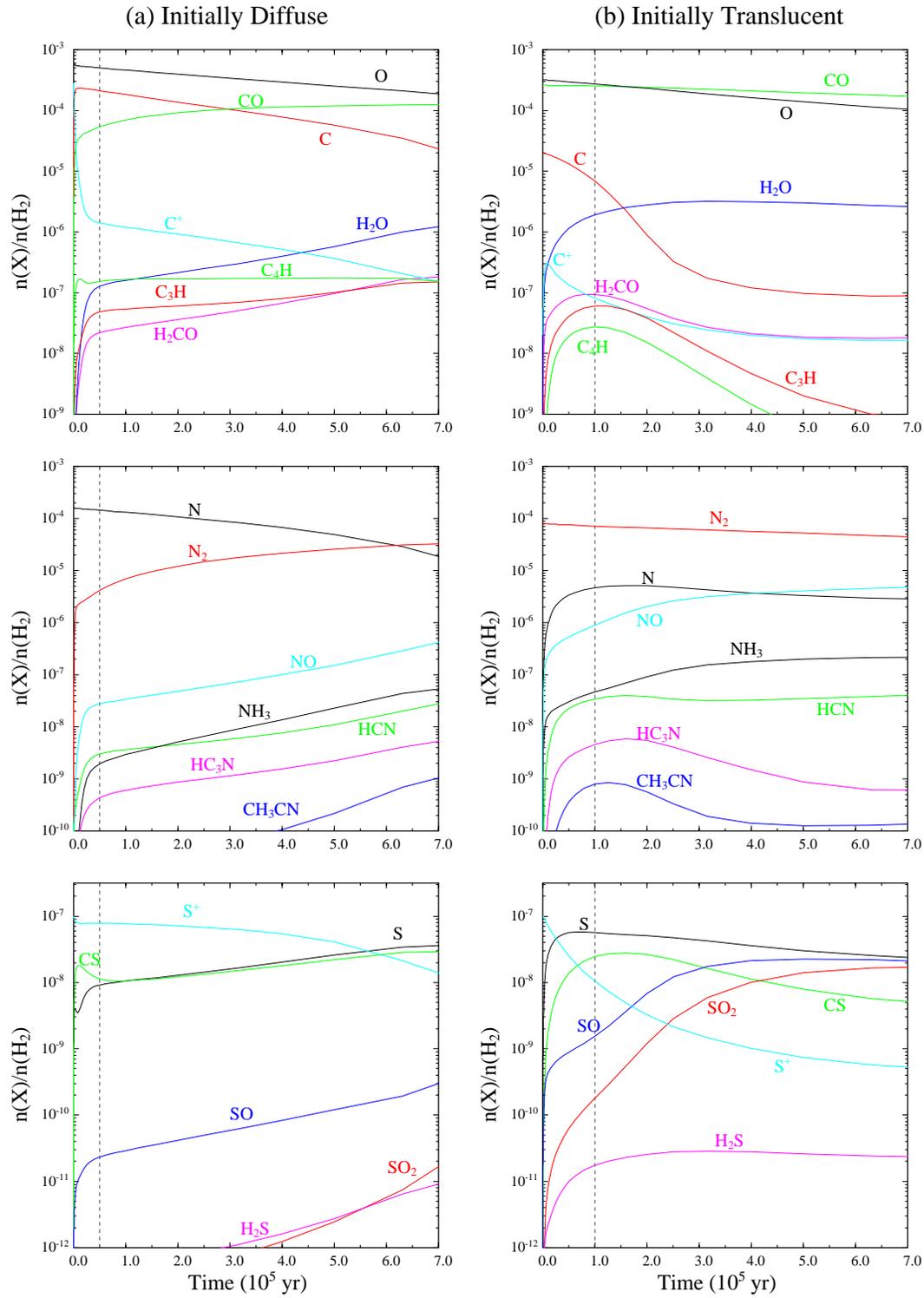}
\caption{As figure 1, but for Globule 2.}
\label{fig:G2}
\end{figure*}

\begin{figure*}
\includegraphics[width=6in]{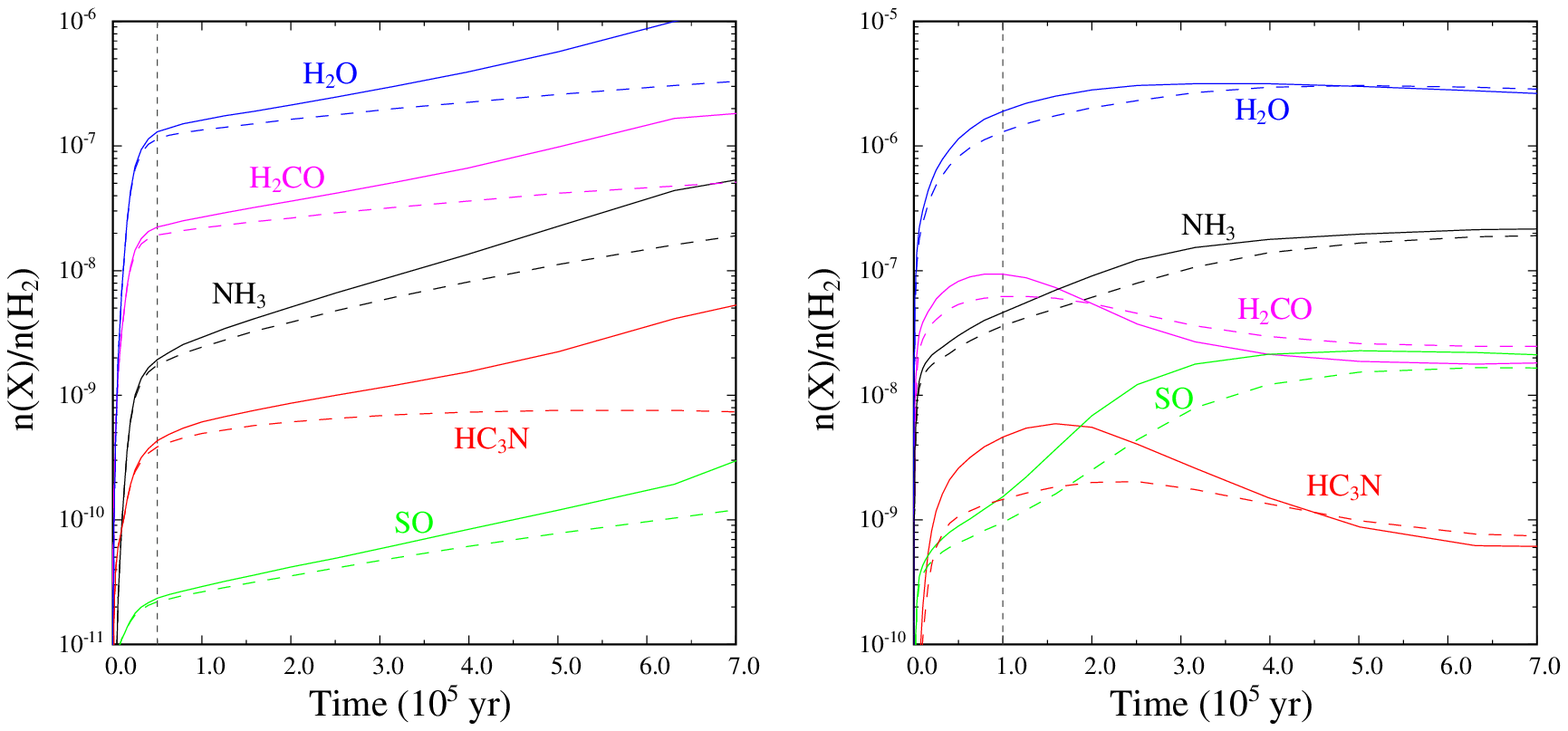}
\caption{Effects of metals on the chemical evolution in
Coalsack Globule 2. As in figures 1 and 2, the left and right panels
show the results for models D and T respectively, and the gray vertical dashed
lines mark the ice mantle formation time-scale, \tm.
Solid lines show the results for models
with zero metals, and dashed lines show the results with
a silicon abundance of 10$^{-7}$.}
\label{fig:metals}
\end{figure*}

Figure~\ref{fig:G1} shows the chemical evolution in Globule 1 for both
the initially diffuse and initially translucent model runs. Also
indicated on the graphs are the mantle formation timescales, \tm,
derived from the H$_2$O ice observations. The values of
\tm\ are different for each model since \tm\ depends on the initial
abundance of atomic O\@.  Figure~\ref{fig:G2} shows the evolution in
the denser Globule 2.
Globule 3 has a density
intermediate to Globules 1 and 2, and the chemical evolution in this
source is similar to that in Globule~1.

\begin{table*}
\caption{Calculated molecular column densities in each of the three globules.}
\label{tab:results}
\begin{center}
\begin{tabular}{lrrrrrr}
\hline\hline
Molecule & \multicolumn{2}{c}{Globule 1} & \multicolumn{2}{c}{Globule 2} &
\multicolumn{2}{c}{Globule 3} \\
& D~~~~ & T~~~~ & D~~~~ & T~~~~ & D~~~~ & T~~~~ \\
\hline \tm ($10^5$~yr) & 2.0~~~ & 3.7~~~ & 0.5~~~ & 1.0~~~ & 0.4~~~ & 0.7~~~ \\
\hline
C         & 4.6(17)  & 4.1(15)  & 7.0(17)  & 6.1(14)  & 1.3(18)  & 1.0(15)  \\
C$^+$     & 1.0(15)  & 1.9(14)  & 1.9(14)  & 9.2(13)  & 4.0(15)  & 1.4(14)  \\
CO        & 1.3(18)  & 7.1(17)  & 1.6(18)  & 1.5(18)  & 8.3(17)  & 1.6(18)  \\
CH$_4$    & 7.6(15)  & 6.8(14)  & 5.3(16)  & 7.0(14)  & 6.1(14)  & 7.8(14)  \\
H$_2$CO   & 4.0(15)  & 4.5(14)  & 1.4(16)  & 1.8(14)  & 6.1(14)  & 2.0(14)  \\
H$_2$O    & 7.4(15)  & 1.1(16)  & 3.5(16)  & 1.9(16)  & 2.7(15)  & 2.1(16)  \\
OH        & 6.6(14)  & 5.4(15)  & 5.3(14)  & 4.0(15)  & 2.1(14)  & 3.2(15)  \\
O$_2$     & 4.6(14)  & 6.2(16)  & 6.2(14)  & 1.7(17)  & 1.5(14)  & 1.1(17)  \\
NH$_3$    & 7.8(14)  & 4.1(15)  & 9.0(14)  & 3.0(15)  & 6.8(13)  & 1.5(15)  \\
NO        & 1.6(15)  & 4.5(16)  & 1.0(15)  & 3.3(16)  & 3.1(14)  & 2.1(16)  \\
HCN       & 6.4(14)  & 1.6(15)  & 4.6(15)  & 4.7(14)  & 1.0(14)  & 2.6(14)  \\
HNC       & 2.9(14)  & 7.8(14)  & 1.2(15)  & 2.7(14)  & 3.4(13)  & 1.6(14)  \\
CN        & 3.6(14)  & 5.8(14)  & 2.7(14)  & 1.2(14)  & 8.9(13)  & 1.2(14)  \\
CS        & 3.8(14)  & 8.8(13)  & 4.8(14)  & 7.4(13)  & 2.4(14)  & 1.3(14)  \\
SO        & 6.7(11)  & 9.9(13)  & 2.4(11)  & 2.5(14)  & 3.3(11)  & 1.5(14)  \\
SO$_2$    & 2.2(10)  & 6.0(13)  & 3.7(10)  & 1.8(14)  & 5.3(9)   & 5.6(13)  \\
H$_2$S    & 1.0(11)  & 4.5(11)  & 1.8(11)  & 6.0(11)  & 1.1(10)  & 3.0(11)  \\
H$_2$CS   & 3.7(13)  & 9.6(11)  & 7.0(13)  & 2.4(12)  & 6.4(12)  & 7.0(12)  \\
C$_2$H    & 5.7(14)  & 2.4(13)  & 3.8(14)  & 5.5(12)  & 1.1(14)  & 7.9(12)  \\
C$_3$H    & 3.9(15)  & 4.6(13)  & 7.6(15)  & 1.6(13)  & 1.1(15)  & 9.2(13)  \\
C$_3$N    & 4.9(15)  & 4.6(12)  & 2.5(16)  & 2.3(12)  & 4.7(15)  & 1.8(13)  \\
C$_4$H    & 4.1(15)  & 1.5(13)  & 8.7(15)  & 4.7(12)  & 2.4(15)  & 3.4(13)  \\
C$_3$H$_2$& 2.1(14)  & 2.9(13)  & 1.2(14)  & 7.7(12)  & 5.4(13)  & 2.1(13)  \\
HC$_3$N   & 2.5(14)  & 2.5(13)  & 3.4(15)  & 9.0(12)  & 2.6(13)  & 2.2(13)  \\
HCO$^+$   & 4.0(13)  & 9.8(13)  & 7.2(13)  & 1.4(14)  & 3.5(12)  & 1.2(14)  \\
O         & 9.7(17)  & 3.9(17)  & 2.3(18)  & 9.0(17)  & 2.8(18)  & 1.3(18)  \\
CH        & 2.4(14)  & 6.6(12)  & 2.7(14)  & 2.6(12)  & 2.8(14)  & 3.0(12)  \\
CH$_2$    & 2.9(13)  & 6.7(11)  & 5.3(13)  & 2.9(11)  & 5.7(13)  & 3.1(11)  \\
C$_2$H$_2$& 1.5(14)  & 5.7(13)  & 1.7(14)  & 5.2(13)  & 3.3(13)  & 1.1(14)  \\
N$_2$     & 4.8(17)  & 6.4(17)  & 4.2(17)  & 8.1(17)  & 1.3(17)  & 6.0(17)  \\
NH        & 2.1(14)  & 1.5(15)  & 2.7(13)  & 2.8(14)  & 9.7(12)  & 1.2(14)  \\
CH$_3$CN  & 6.2(13)  & 1.3(13)  & 1.4(15)  & 1.6(12)  & 1.7(12)  & 1.5(12)  \\
CH$_3$OH  & 9.6(9)   & 4.6(10)  & 1.7(11)  & :.0(10)  & 1.1(8)   & 1.5(11)  \\
CH$_3$OCH$_3$ & 6.3(6)   & 4.2(6)   & 3.6(8)   & 6.0(6)   & 1.2(4)   & 8.3(6)   \\
OCS     & 1.3(12)  & 4.1(11)  & 2.4(12)  & 1.7(12)  & 1.8(11)  & 1.9(12)  \\
N$_2$H$^+$ & 9.7(12)  & 5.4(13)  & 7.0(12)  & 3.0(13)  & 3.3(11)  & 1.8(13)  \\

\hline\hline
\end{tabular} \\
\end{center}
\begin{flushleft}
{\small a(b) represents $a\times10^{b} ~\rm cm^{-2}$. \\ Column
  densities are calculated by multiplying the abundance calculated by
  our chemical model at time \tm\ by the H$_2$ column densities
  determined from the visual extinction measurements of Smith et
  al. (2002). \\ D and T refer to models
  with the initial conditions appropriate for diffuse or translucent
  material respectively (see table~\protect{\ref{tab:init}}).   }
\end{flushleft}
\end{table*}

Comparing figures \ref{fig:G1} and \ref{fig:G2} it is apparent that  
although the chemical evolution of both globules is similar in many respects,
the difference in density results in a slightly different chemistry in
each source. If the globules formed rapidly from diffuse cloud material,
it takes a long time for the conversion of the
initially atomic material into predominantly molecular form, but in
the denser Globule 2 the CO/C ratio becomes $>1$ after
$3\times10^5$~yr, whereas in the lower density Globule 1 the
transition takes almost 1~Myr. The transformation of atomic nitrogen
into N$_2$ takes even longer in each source. The large abundances of
C$^0$ and N$^0$ act to suppress the abundances of species such as CN,
NO, and SO\@, which in turn tend to reduce the abundances of many
other species below their `typical' dark cloud values. In both globules
HCN, NH$_3$, HC$_3$N, and CH$_3$CN remain at low abundances for $t <
1$~Myr, although their abundances become somewhat larger in Globule 2
at later times. Based on our values of \tm, Globule 1 is older than Globule 2, and so the
difference in abundances between the two regions will not be as
pronounced as if both were the same age, but we predict that in
Globule 2 the abundances of H$_2$CO, HCN, HC$_3$N, and CS,
should be $\approx 2$--3 times larger than in Globule 1.

The alternative scenario, in which CO and N$_2$ formed prior to the
cores reaching the critical extinction for water ice formation to
occur, results in larger abundances of several molecules, especially
sulphur-bearing species. For example, SO, SO$_2$, and H$_2$S are much
more abundant in panels (b) of figures \ref{fig:G1} and \ref{fig:G2}. Again, chemical
differences between the two sources arise primarily due to their
different ages, and we expect that carbon chain species such as C$_3$H and C$_4$H
should be 2--3 times more abundant in the younger Globule 2, whereas
NH$_3$, SO and SO$_2$ should be more abundant in Globule 1.

In terms of discriminating between the two model scenarios, there are
large differences in the abundances of several species. If the globules
formed directly from diffuse material, the abundances of carbon-chain
molecules such as C$_4$H will be larger than if the gas was initially
molecular. On the other hand, the reduced C$^0$ abundance in the
latter case leads to larger abundances of many other molecules such as
H$_2$CO, NH$_3$, HCN, HNC, and especially SO and SO$_2$\@. Therefore,
observations of these species in the Coalsack can in principle be used
to determine which model is correct.  In order to allow comparison
with observations, table~\ref{tab:results} lists the column densities
for selected species in each of the three globules at the appropriate
time, \tm, calculated for each model.

\subsection{Effects of model parameters}
\label{sec:metals}

We have also investigated the effects of altering various parameters
of our models. As discussed above, we used the elemental abundances of
Savage \& Sembach (1996). However, if we consider the only direct measurement
taken toward the Coalsack, Jensen, Rachford \& Snow
(2005) reported a low elemental O/H ratio of only $6.5\times10^{-5}$
toward HD 110432, a factor of four less than used in our model (although
the error bars on this measurement are also consistent with a `normal' O/H ratio.) 
We therefore ran our model with the abundances of all elements depleted
by a factor of four. In the case of initially translucent conditions (model T) the
effects are relatively minor, but for model D we find that the reduced elemental
abundances act to speed up the transition from the diffuse-cloud-type chemistry to the
dark-cloud-type chemistry, resulting in enhanced abundances of
species such as H$_2$CO, NH$_3$, HCN, SO, and SO$_2$\@. However, a key
aspect of this change is that it also increases the values of \tm\ by a factor of four,
since the O atom freeze-out rate is reduced. For the observed ice thicknesses
in Globule 1 this implies that that this source must have an age comparable to or
greater than the accretion time-scale, which is equal to 1.2~Myr. This
conflicts with other evidence that the \SC\ is a relatively young
region (see $\S$~2), and so we consider it likely that the elemental
abundances (or at least the oxygen abundance) in the Coalsack are
close to standard interstellar values.

We also investigated the effects of a non-zero metal abundance, by running each model
with an initial abundance of Si$^+$ equal to $10^{-7}$.  We find that
the effects on the chemistry of Globule 1 are minor, but in the denser
Globule 2 the effects are more important. Figure~\ref{fig:metals} compares
the abundances of those species most affected by the metal abundance in
models D and T for Globule 2. We see that the effects are largest in model D at
late times; at the early times implied by the observed water ice
column densities the presence or absence of metals has little influence on
the predicted abundances. One exception is HC$_3$N, where the early-time abundance
is depressed in model T when Si$^+$ is included.

\section{Discussion}

Most models of interstellar dark cloud chemistry assume initial
conditions appropriate for diffuse clouds, i.e.\ they assume that the
cloud formation timescale is much less than the chemical timescale.
For typical dark cloud densities of $n({\rm H}_2) = 10^4$~cm$^{-3}$,
the exact choice of the initial conditions is not so important, as the
chemistry rapidly evolves from the diffuse cloud conditions into
`standard' dense cloud chemistry. Our results show that, for lower
density gas such as the globules in the Coalsack, the transformation
time may be much longer, of order Myr. In this case, the time spent in
the translucent phase is a key parameter of the model, and the dense
globules retain a chemical `memory' of the prior conditions. Hence, in
principal, molecular observations of the dark globules can be used to
constrain the dynamics of their origins.

Rachford et al.\ (2001) observed CO and H$_2$ toward the field star HD 110432, and
derived a CO/H$_2$ column density ratio of $\approx 3\times10^{-7}$,
suggesting that the molecular fraction in the inter-clump medium is
low. However, at such low H$_2$ column densities the overall CO/H$_2$
ratio may be low due to CO dissociation in the outer layers where
$A_{\rm V} \ltsim 1$, and is still consisitent with almost all C in CO
in the central regions (e.g.\ van Dishoeck \& Black 1989). It is
difficult to use such observations to tightly constrain the C/CO ratio
in the interclump medium.

Although there have been several studies of dust and extinction in the
Coalsack, there is a dearth of molecular line observations of this
region. As far as we know, the only molecules observed in the Coalsack
are CO (Huggins et al.\ 1977; Codina et al.\ 1984; Kato et al.\ 1999),
OH, and H$_2$CO (Brooks et al.\ 1976). The latter observed the 6~cm
formaldehyde $1_{11}$--$1_{10}$ line in absorption, 
and derived a H$_2$CO/H$_2$ abundance of $\sim$ 10$^{-9}$,
at least an order of magnitude smaller than the abundances we derive for the globules in 
Table 3.  However, the 6~cm line absorption feature appears to arise preferentially from
the lower density regions surrounding the dense core rather than from the dense core itself 
(Snell 1981).  This would appear to be confirmed by the fact that Brooks et al.\ report 
the emission from the H$_2$CO is extended over a size of 12 arcminutes 
(similar to that of the OH emission), and also by the fact that our chemical
model predicts a H$_2$CO abundance of $\sim 10^{-9}$ in low-density, translucent
material. A second effect to consider is that extinction maps reveal the densest regions of 
the globules are likely of order 1--3 arcminutes in size (Bok 1977; Jones et al.\ 1980;
Lada et al.\ 2004), 
smaller than the 4 arcminute beamsize of Brooks et al. 
Hence, H$_2$CO features arising from the centers of the globules will
suffer from beam dilution. We conclude that 
the 6~cm observations of Brooks et al.\ most likely trace the interclump 
medium.

\begin{figure} 
\includegraphics[width=8.5cm]{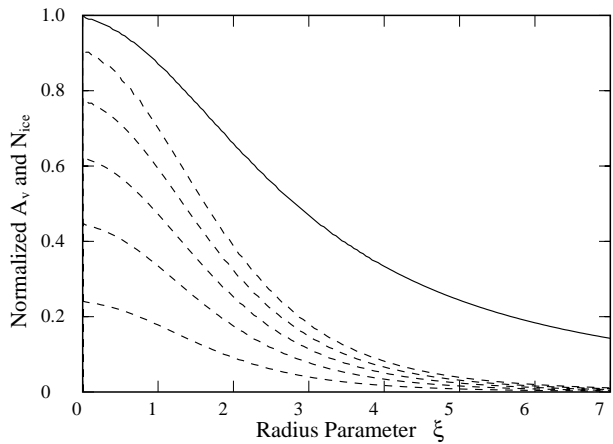}   
\caption{Integrated pencil-beam H$_2$ column density (i.e.\ $A_{\rm
    v}$; solid line) and water ice column densites (dashed lines) as a
  function of impact parameter for a Bonnor-Ebert sphere with central
  density $n_{\rm H} = 10^4$~cm$^{-3}$, appropriate for Globule 2 in
  the Coalsack. Ice column densities are shown at times of 2, 4, 6, 8, \&
  10 $\times$ $10^5$~yr.}
\label{fig:benice} 
\end{figure}


We have only considered the case of a single sightline through a
constant density medium since that is the nature of the observational
constraints currently available to us (Smith et al.\ 2002).  In
reality, such prestellar globules have spatial density gradients, and
the observed ice column density depends on the particular
line-of-sight path integral through the cloud. Recent studies of the
infrared extinction of field stars which lay behind these globules
have allowed detailed maps to be made of the visual extinction through
several globules (Alves, Lada \& Lada 2001; Kandori et al.\ 2005).  For
isolated globules, the dust density profile is usually well matched by
a Bonnor-Ebert sphere (Alves et al.\ 2001; Racca et al.\ 2002) and many
appear to be on the critical limit of gravitational instability
against runaway collapse (Kandori et al.\ 2005).  The ice formation
time-scale derived for any sightline will always lag that of the
dynamical time-scale over which the mass distribution was
assembled. Hence, if 3 micron observations towards fields stars behind
such globules can be made to allow their {\it radial ice profile} to
be constructed, this could eventually be employed as a good measure of
the dynamical time.  The analysis presented in this paper allows us to
calculate the water ice column density as a function of both time and
offset, $\xi$, from the core center in Coalsack Globule 2, where
$n_{\rm H} (r, \xi) $ is from a Bonnor-Ebert sphere (Racca et
al.\ 2002). As an example, Figure~(\ref{fig:benice}) shows that the ice
profile has a distinctive shape at each time, and so fitting the
observed $N_{\rm ice} (\xi) $ distribution with such models will allow
a global derivation of \tm\ for each source.

\section{Conclusions}

Oxygen atom accretion and reaction to
form H$_2$O ice is a `one-way' process (at least until a protostar has
formed), is essentially unaffected by gas-phase chemistry, and begins
only when the dense core forms. Hence, ice column densities can be  
employed as an independent measure of physical age.  A combination of ice absorption and
molecular emission observations can thus  lead to quantitative measures of
globule age, and allow correlations between physical and chemical
evolution to be explored.
 
We have determined the `ice mantle age', \tm, of three dense
globules in the Southern Coalsack, based on the time needed to deposit
the H$_2$O ice abundances observed by Smith et al.\ (2002) in these
sources. In conjunction with a simple chemical model, we have used
these values of \tm\ to predict the abundances of numerous gas-phase
molecules toward each of these sources. These predictions can be
easily tested by future molecular  line observations of the Coalsack. Our main
findings can be summarized as follows:

\begin{itemize}
\item In each globule only a small fraction of the total available gas-phase
 oxygen is frozen out as water, indicating that each must be significantly
 younger than the accretion time-scale.
\item We derive values for \tm\ of a few times $10^5$~yr.\ for Globule
  1 and $\la 10^5$~yr.\ for Globules 2 and 3. These ages are
  consistent with the dynamical age for Globule 2 derived by Lada et
  al.\ (2004).
\item The principal uncertainty in determining exact values for
  \tm\ arises from our lack of knowledge of the gas-phase O atom
  abundance in each source and the grain size distribution.
  However, assuming each globule formed from
  chemically similar gas, the {\it relative} ages of the globules are
  easily constrained.
\item The chemical evolution is very sensitive to the initial
  conditions when the visual extinction becomes large enough for water
  ice to accrete. In particular, whether carbon and nitrogen are
  atomic or locked up in CO and N$_2$ has a profound influence on the
  chemistry: if ${\rm C/CO > 1}$ at $t = 0$, the abundances of many
  species are less than if the material is initially in molecular
  form.
\item As the globules age, the abundances of some species change, and
  should be different in each of the three globules. Observations of
  several species in each of the three globules can be used to test
  the validity of our model.
\item The relative abundances of sulphur-bearing species,  particularly SO and H$_2$S, 
appear to be  useful diagnostics of the differences between the various models.
\item Combining water ice observations with gas-phase molecular line observations 
toward pre-stellar cores provides a method of dating these objects. Such measurements
can potentially discriminate between `fast' and `slow' scenarios for molecular cloud formation.
 \end{itemize} 


\section*{Acknowledgments}  
This work was supported by NASA's Long Term Space Astrophysics
Program, with partial support from the Origins of Solar Systems
Program, through NASA Ames Cooperative Agreement NCC2-1412 with the
SETI Institute.



\end{document}